\documentclass[aoas]{imsart}

\usepackage{xr}
\usepackage{xr-hyper}

\RequirePackage[OT1]{fontenc}
\usepackage{amsthm,amsmath,natbib,algorithm,algorithmic}
\usepackage{amssymb}
\RequirePackage[colorlinks,citecolor=blue,urlcolor=blue]{hyperref}
\RequirePackage{natbib}
\usepackage{hyperref}
\usepackage{url}
\usepackage{subfigure}
\usepackage{float}
\usepackage{verbatim}
\usepackage{graphicx}

\usepackage{bm,bbm}
\newcommand{\E}{\mathbb{E}}

\startlocaldefs
\numberwithin{equation}{section}
\theoremstyle{plain}

\newtheorem{prop}{Proposition}[section]
\endlocaldefs

\begin{document}

\begin{frontmatter}
\title{Nonparametric Bayesian multi-armed bandits for single cell experiment design}
\runtitle{Bayesian nonparametric experimental design}
%\thankstext{T1}{Footnote to the title with the `thankstext' command.}

\begin{aug}
\author{\fnms{Federico} \snm{Camerlenghi}\thanksref{t1,t3}\ead[label=e1]{federico.camerlenghi@unimib.it}},
\author{\fnms{Bianca} \snm{Dumitrascu}\thanksref{t1}\ead[label=e2]{biancad@ias.edu}},
\author{\fnms{Federico} \snm{Ferrari}\thanksref{t1}\ead[label=e3]{ff31@duke.edu}},
\author{\fnms{Barbara E.} \snm{Engelhardt}\ead[label=e4]{bee@princeton.edu }}
\and
\author{\fnms{Stefano} \snm{Favaro}\thanksref{t2}\ead[label=e5]{stefano.favaro@unito.it}}
%\ead[label=e3]{third@somewhere.com}
%\ead[label=u1,url]{www.foo.com}}

\thankstext{t1}{These authors contributed equally to this work.}
\thankstext{t3}{Also affiliated to Collegio Carlo Alberto, Torino and BIDSA, Bocconi University, Milano, Italy.}
\thankstext{t2}{Also affiliated to IMATI-CNR ``Enrico  Magenes" (Milan, Italy).}
%\thankstext{t4}{Part of this work was completed during the author's visit with SAMSI and the Department of Statistical Science, Duke University, Durham, NC, USA.}

\runauthor{Camerlenghi, Dumitrascu, Ferrari et al.}

\affiliation{
University of Milano - Bicocca, Institute for Advanced Study,  Duke University, Princeton University,  University of Torino and Collegio Carlo Alberto}

\address{Federico Camerlenghi\\
Department of Economics, Management and Statistics\\
University of Milano - Bicocca\\
 20126 Milano, Italy\\
\printead{e1}}

\address{Bianca Dumitrascu\\
School of Mathematics\\
Institute for Advanced Study\\
1 Einstein Drive, Princeton, NJ, 08540 USA \\
\printead{e2}}

\address{Federico Ferrari\\
Department of Statistical Science\\
Duke University\\
415 Chapel Dr, Durham, NC 27705 \\
\printead{e3}}

\address{Barbara E. Engelhardt\\
Department of Computer Science\\
Center for Statistics and Machine Learning \\
Princeton University\\
35 Olden Street, Princeton, NJ 08540 \\
\printead{e4}}

\address{Stefano Favaro\\
Department of Economics and Statistics\\
University of Torino\\
10134 Torino, Italy\\
\printead{e5}}

\end{aug}

\begin{abstract}
The problem of maximizing cell type discovery under budget constraints is a fundamental challenge for the collection and analysis of single-cell RNA-sequencing (scRNA-seq) data. In this paper, we introduce a simple, computationally efficient, and scalable Bayesian nonparametric sequential approach to optimize the budget allocation when designing a large scale experiment for the collection of scRNA-seq data for the purpose of, but not limited to, creating cell atlases. Our approach relies on the following tools: i) a hierarchical Pitman-Yor prior that recapitulates biological assumptions regarding cellular differentiation, and ii) a Thompson sampling multi-armed bandit strategy that balances exploitation and exploration to prioritize experiments across a sequence of trials. Posterior inference is performed by using a sequential Monte Carlo approach, which allows us to fully exploit the sequential nature of our species sampling problem. We empirically show that our approach outperforms state-of-the-art methods and achieves near-Oracle performance on simulated and scRNA-seq data alike.  HPY-TS code is available at \url{https://github.com/fedfer/HPYsinglecell}.
\end{abstract}

\begin{keyword}
\kwd{cell type discovery}
\kwd{experimental sampling design}
\kwd{hierarchical Pitman-Yor model}
\kwd{multi-armed bandits}
\kwd{scRNA-seq}
\kwd{sequential Monte Carlo}
\kwd{Thompson sampling}
\end{keyword}

\end{frontmatter}

%%%%%%%%%%%%%%%%%%%%%%%%%%%%%%%%
%%%%%%%%%%%%%%%%%%%%%%%%%%%%%%%%
%%%%%%%%%%%%%%%%%%%%%%%%%%%%%%%%
%%%%%%%%%%%%%%%%%%%%%%%%%%%%%%%%
\section{Introduction}

Technological developments in high-throughput genomics have generated a wealth of data allowing researchers to measure and quantify RNA levels of individual cells~\citep{single_cell_1,single_cell_2}. Benefiting from experimental and computational advances alike, single-cell RNA-seq (scRNA-seq) allows the characterization of cell types and cellular diversity, offering invaluable insights at scales unattainable in previous bulk gene expression studies \citep{bulk_v_sc_1}. In order to understand the diversity of the thousands of cell types and subtypes across different organisms, recent initiatives aim for molecular profiling of all cell types of complex organisms such as mouse or human \citep{regev2017science, han2018mapping}. Despite the decreasing cost of technologies for single-cell sequencing, cell atlases are expensive to collect and hard to coordinate across species, cells, tissues, organs, diseases, technologies, and labs. A principled way of collecting data is therefore paramount: given the experimental cost limiting the number of cells to be sequenced, and given multiple related experimental scenarios (e.g., developmental time, biological region, tumor site), how can one allocate the cellular sequencing budget in order to minimize experimental cost and to maximize the number of distinct cells types obtained? In this paper, we present an effective Bayesian nonparametric approach to address this fundamental question. 

Recent work \citep{bubeck2013optimal, battiston2016multi, dumitrascu2018gt} proposed the use of classical multi-armed bandit strategies---upper confidence bounds (UCB) \citep{lai1985,auer2002} and Thompson sampling (TS) \citep{thompson1933}---for devising sequential approaches to maximize the number of distinct species discovered by sampling over multiple populations. In particular, these sampling strategies balance the exploration of the experimental choices---which populations are sampled---with the exploitation of populations that maximize current estimates of the expected rewards--the observed species diversity within a population. In the classical multi-armed bandit setting, a gambler is presented with slot machines (\emph{one-armed bandits} is the colloquial term for a slot machine in American slang) that each pay out a random reward sampled from an arm-specific probability distribution. The gambler commits to querying a given arm for a single trial before switching to another arm, and her goal is to select a sequence of arms to play in order to maximize her rewards over subsequent trials. At each step, the gambler estimates the expected rewards of a single trial from each machine's arm, both queried and not. She must then balance exploiting the arm with the current highest estimate of expected rewards and exploring undersampled arms to improve estimates of the arms' expected rewards. 

A natural variation of the above multi-armed bandit setting is when the gambler commits to querying a given arm for a pre-determined number of consecutive trials before switching to another arm. This variation is readily applicable to the experimental design problem of guiding the sequential selection of samples through single cell sequencing technologies: We may sequence some number of cells from one of multiple tissues or sample sites. In detail, we consider this problem as a set of sequential trials where a scientist may choose a subset of tissue samples to assay. Each organ, tissue type, sample site, or experimental condition represents an arm to be pulled. When choosing a specific arm, the scientist commits to sequencing a number of cells proportional to the maximum number of new cell type discoveries expected in a future sample from the given experimental condition. The reward of each experimental trial is given by the number of new cell types uncovered in the sequenced sample. A first attempt to address this problem, within the context of scRNA-seq data, was proposed in \cite{dumitrascu2018} by combining a class of Good-Toulmin (GT) estimators \citep{good1953population, good1956number,efron1976estimating, orlitsky2016optimal} with the TS strategy.

In this paper, we follow ideas from \cite{battiston2016multi} and \cite{dumitrascu2018} to introduce a Bayesian nonparametric counterpart of the previous Good-Toulmin Thompson sampling (GT-TS) approach \citep{dumitrascu2018}. Because of the purely nonparametric nature of smoothed GT estimators, the GT-TS approach does not allow us to take into account the structure of cell type diversity. As cell types arise through \emph{cellular differentiation} \citep{rizvi2017single}, they organize themselves in developmental landscapes \citep{waddington1957strategy}. Hierarchical structures can be imposed on the cell types through Bayesian nonparametric priors, as was done for cell trajectory reconstruction and Bayesian inference on developmental lineages \citep{heaukulani2014beta, shiffman2018reconstructing}. 

A natural choice for a nonparametric prior to model cell type diversity is the hierarchical Pitman-Yor process (HPY)~\citep{teh2006hierarchical,teh2010}. The HPY process has previously been used in the context of species discovery problems in multiple populations, and it has been shown to have good performance in small data sets \citep{camerlenghi2018,bassetti2018}. Yet, species sampling problems considered in these recent studies are not sequential problems: a Bayesian nonparametric model with a HPY prior is fit to the data \emph{de novo} each time new data become available. This makes current posterior sampling procedures designed for the HPY prior infeasible for our sequential species sampling problem of rapidly-growing single cell data sets. 

We propose a simple, computationally efficient, and scalable Bayesian nonparametric sequential approach for guiding the selection of samples for single cell sequencing technologies with the goal of maximizing the diversity of cell types discovered. Our approach has two main contributions. First, we introduce a multi-armed bandit strategy that combines the TS approach with a Bayesian nonparametric counterpart of the GT estimator under the HPY prior, extending previous work that allowed only a single trial to allow a pre-determined number of consecutive trials before switching arms~\citep{battiston2016multi}. The TS strategy encodes the sequential exploration-exploitation process associated with data collection from any given region, whereas the use of the HPY prior incorporates biologically-relevant information regarding the relationships among cell types to guide the allocation of resources. Second, we devise an efficient posterior sampling scheme that relies on sequential Monte Carlo methods \citep{west1993a, liu2001combined}. Sequential Monte Carlo (SMC) allows us to fully exploit the sequential nature of our species sampling problem, thus avoiding the overwhelming computational burden of the Markov chain Monte Carlo (MCMC) scheme proposed in \cite{battiston2016multi}.
We compare our method to the previous method (GT-TS) and to an Oracle in simulations and in a data set based on the Mouse Cell Atlas \citep{han2018mapping}. Since our motivation lies in the realm of single cell experimental design, we illustrate how, given a per-trial budget, the resulting algorithm leverages information across tissues to inform subsequent experiments in order to maximize cell type discovery in the Mouse Cell Atlas \citep{han2018mapping}.

The paper is structured as follows. Section \ref{sec:preliminary} contains preliminaries on i) the multi-armed bandit setting within the context of prioritizing single cell sampling across populations, i.e., organs, tissues, regions, and experimental conditions; ii) the definition of the HPY prior, and some of its marginal sampling properties. In Section \ref{sec:inference}, we introduce our Bayesian nonparametric sequential approach, referred to as the HPY-TS strategy, for guiding the selection of samples through single cell sequencing technologies. A detailed description of the sequential Monte Carlo approach for posterior sampling is presented in Section \ref{sec:algo}. Section \ref{sec:application} highlights results of our approach through a simulation study and an application to a data set derived from the Mouse Cell Atlas. In Section \ref{sec:discussion}, we summarize our work and briefly discuss extensions to our HPY-TS strategy. Additional simulation studies, posterior diagnostics, and proofs are deferred to the Supplementary Material \citep{suppm}.

%%%%%%%%%%%%%%%%%%%%%%%%%%%%%%%%
%%%%%%%%%%%%%%%%%%%%%%%%%%%%%%%%
%%%%%%%%%%%%%%%%%%%%%%%%%%%%%%%%
%%%%%%%%%%%%%%%%%%%%%%%%%%%%%%%%
\section{Preliminaries}  \label{sec:preliminary}

Let $\mathcal{Y}$ denote the set of labels representing the cell types of an organism being studied. The cell type composition within each of the $J$ possible populations (arms, experiments) is characterized by a probability distribution over $\mathcal{Y}$, such that cell types are shared across the $J$ populations. Precisely, we denote by $P_j$ the probability distribution on $\mathcal{Y}$ in population $j$, for $j=1,\dots,J$. Let $n_j$ be the number of cells (pulls) observed from the $j$th population, let $\bm{Y}_{j} = (Y_{j,1},\cdots,Y_{j,n_j}) \in \mathcal{Y}^{n_j}$ be the vector of $n_j$ observations from the $j$th population, whereas $\bm{Y} = (\bm{Y}_{1},\cdots,\bm{Y}_{J})$ is the joint sample corresponding to a budget $n_1+ \cdots +n_J$. Assume now to have an additional budget $M$ that constraints the number of cells that can be collected per trial in a future experiment. If $J$ populations are available, a multi-armed bandit iteratively selects a subset of populations to sample from as well as the appropriate number of cells to sample in each population. In other words, at each step, the arm is chosen with the goal of maximizing the number of novel cell types observed. Therefore, in order to set up our strategy, the first step consists in estimating the number of thus far unseen species (cells) that are going to be sampled for every possible arm $j$, as $j=1, \ldots, J$.

A possible strategy to address this sequential problem was first proposed in the work of \citet{dumitrascu2018gt}. This approach relies on a smoothed version of the Good-Toulmin estimator of the number of unseen species~\citep{orlitsky2016optimal}. However, while the smoothed Good-Toulmin estimator presents attractive statistical properties and provable guarantees in terms of minimax optimality \citep{orlitsky2016optimal}, it is designed for a single population scenario. In this paper, we focus on data coming from multiple related populations. Indeed, we discover cell types across diverse tissue types assayed in scRNA-seq experiments. It is then important to guarantee two key properties in our model: i) the model preserves data heterogeneity for different tissues; and ii) the model allows borrowing of information across the different tissues. Hierarchical Bayesian nonparametric priors are tailored for such situations: the data are divided into distinct populations (according to the tissue they are derived from), and at the same time the hierarchical construction allows a borrowing of  information across the diverse populations of cell types.

\subsection{The Pitman-Yor process}

The Bayesian nonparametric (BNP) approach relies on the choice of a prior distribution for the cell type labels. The Dirichlet process (DP) \citep{ferguson1973bayesian} is a well-known Bayesian nonparametric distribution. In this paper, we make use of a generalization of the DP, the Pitman-Yor (PY) process \citep{pitman1997}. The PY process $P$ is a random probability measure that depends on two parameters $(\sigma, \theta )$, respectively called the \emph{concentration} and the \emph{mass} parameter, with a base measure $P_0$ on the space of labels $\mathcal{Y}$. The admissible values we consider here for these parameters are $\sigma \in (0,1)$ and 
$\theta >0$. The most simple way to define the PY process uses a stick-breaking procedure~\citep{sethuraman1994constructive}. More specifically, $P$ is a discrete random probability measure $P=\sum_{k \geq 1} \pi_k \delta_{y_k}$ such that
\begin{displaymath}
\pi_1 = V_1 
\end{displaymath}
and 
\begin{displaymath}
\pi_k = V_k \prod_{h=1}^{k-1} (1-V_h) , \; \text{for } h \geq 2,
\end{displaymath}
where  $(y_k)_{k \geq 1}$ is a sequence of i.i.d. random variables as $P_0$, and $(V_k)_{k\geq1}$ is a collection of independent beta-distributed random variables with parameters $(\theta + k \sigma , 1-\sigma)$. The two sequences $(y_k)_{k \geq 1}$ and $(V_k)_{k \geq 1}$ are assumed to be independent. We write 
$P \sim PY (\sigma, \theta ; P_0)$ to denote the distribution of $P$. The classical DP prior can be found as a limiting case of the PY process, letting $\sigma \to 0$.

It is worthwhile to highlight the differences between the PY process and the DP with respect to the predictive distributions they induce. In both cases, the predictive distribution may be represented in terms of the celebrated Chinese restaurant process (CRP); see \cite{pitman1997} for a comprehensive account and references. Consider a random sample $Y_1, \ldots, Y_n | P 
\stackrel{iid}{\sim} P$ of size $n$ from the PY process. The almost sure discreteness of the random probability measure $P$ allows for ties within the sample. Then, let $K_{n}$ be the number of distinct values within the sample $ (Y_1, \ldots , Y_n)$, denoted as $(Y_{1}^{*},\dots,Y_{K_{n}}^{*})$ and having multiplicities $(n_{1}, \ldots, n_{{K_n}})$. Then, the predictive distribution of the $n+1$st observation $Y_{n+1}$ given past observations is 
\begin{equation} \label{eq:prediction}
Y_{n+1} |(Y_1, \ldots , Y_n),P_0,\sigma,\theta \sim \sum_{k=1}^{K_{n}} \frac{n_{k}-\sigma}{\theta+n}\delta_{Y_{k}^{*}}+\frac{\theta
+K_{n} \sigma}{\theta + n}P_0.
\end{equation}
In other words, Equation \eqref{eq:prediction} tells us that the probability of observing a previously observed value $Y_{k}^{*}$ is proportional to $n_{k}-\sigma$. Intuitively, the more samples of a species we observe, the higher the probability of sampling it again in future trials; this is referred to as ``the rich get richer'' behavior. Alternatively, the probability of sampling a new observation from the base measure $P_{0}$ is proportional to $\theta+K_{n} \sigma$. Notice that the clustering structure of the PY process depends on two parameters, $\sigma$ and $\theta$, whereas in the DP it is governed only by $\theta$. This more complex parametrization offers more flexible clustering rates and cluster size tail behaviors for the PY process~\citep{ishwaran2001gibbs}.

\subsection{Hierarchies of Pitman-Yor processes} 

When considering observations sampled from multiple populations, it is natural in the Bayesian framework to model the group structure with a hierarchical framework. Here we use a hierarchical structure based on the PY process.
We denote the distribution of cell type labels across all of the populations (experimental design conditions) by $P$. The probability distribution $P$ is almost surely discrete with an unknown number of atoms, and we select a PY process prior to model the distribution of cell type labels with parameters $(\sigma, \theta)$ and non-atomic base measure $P_0$ on the space of labels. Then each population-specific distribution $P_j$ is modeled using a PY process prior with parameters $(\sigma_j,\theta_j )$, and we further suppose that the common base measure for all the $P_j$s is the PY process $P$. Summing up, we have specified the following hierarchical PY (HPY) process prior:
\begin{gather} \label{eq:hPY_def}
\begin{split}
P |\sigma, \theta, P_0 & \sim PY(\sigma,\theta; P_0)\\[0.2cm]
P_j | \sigma_j,\theta_j,P &\stackrel{ind}{\sim} PY(\sigma_j,\theta_j; P)  \ \ \ \forall j=1,2,\dots,J \\[0.2cm]
Y_{j,i} | P_j &\stackrel{iid}{\sim} P_j  \ \ \ \ \forall j=1,2,\dots,J ,\ \ \ \forall i=1,2,\dots,n_j.
\end{split}
\end{gather}
The above hierarchical specification introduces dependencies among different populations (experimental conditions or arms), thus allowing the sharing of information across populations since the base measure $P$ is common to the different collections of observations \citep{teh2006hierarchical,camerlenghi2018}. In particular, conditional on the \emph{base measure} $P$, the $P_j$s are independent PY processes with base measure $P$. In particular, the interpretation of the parameters $(\sigma_j,\theta_j)$ is the same as in the single population case described above.

The predictive distribution and the combinatorial structure induced by hierarchical processes can be thought of in terms of the Chinese restaurant franchise (CRF) metaphor \citep{teh2010}. 
According to this culinary metaphor, each sample $\bm{Y}_j: = (Y_{j,1}, \ldots , Y_{j,n_j})$ % actually i dont think this part is right but i dont know what y represents.
% It is important to describe the CRF metaphor in terms of tables and restaurants, since this is well known in th whole BNP literature
identifies the dishes chosen by the $n_j$ customers of restaurant (group) $j$, for any $j=1, \ldots, J$. People sitting at the same table eat the same dish, and the same dish can be served within the same restaurant or across different restaurants, since we use the same a.s. discrete base measure $P$ for all of the groups. We denote by $Y_1^{**}, \ldots , Y_{K}^{**} $ the $K$ distinct dishes across the $J$ samples, whereas  $n_{j,k} \geq 0$ represents the number of customers in restaurant $j$ eating dish $k$. Finally, the vector
$\bm{n}_j :=(n_{j,1}, \ldots , n_{j,n_j})$ encodes all of the frequencies for a specific population $j$.

The combinatorial structure induced by the HPY process is formally described by the so called \textit{partially exchangeable partition probability function} (pEPPF) defined by
\begin{equation}
    \label{eq:pEPPF}
    \Pi_k^{(n)} (\bm{n}_1, \ldots , \bm{n}_J) :=
  \E   \int_{\mathcal{Y}^K} \prod_{j=1}^J \prod_{k=1}^{K}
  P_j^{n_{j,k}} ({\rm d} Y_{k}^{**}).
\end{equation}
In other words, this is the probability of observing a specific configuration of the dishes across the restaurants. A tractable expression of the pEPPF \eqref{eq:pEPPF} was found in \citet{camerlenghi2018}, resorting to auxiliary latent variables, which may be  seen as tables in the CRF language. More specifically, each observation (customer) is associated with a latent tag identifying the table of the restaurant at which the specific customer is seated. We have the constraints
 \[
 n_{j,k} = \sum_{t=1}^{m_{j,k}} n_{j,t,k},
 \]
where $m_{j, k}$ is the number of tables in restaurant $j$ serving dish $k$, i.e., $Y_k^{**}$, and $n_{j,t,k}$ is the number of customers in restaurant $j$ sitting at table $t$, eating dish $k$. In the sequel it will be useful to denote by
$K_{j}$ the number of distinct values in the $j$th group $\bm{Y}_j$, indicated by $(Y_{j,1}^{*},\dots,Y_{j,K_j}^{*})$, which is a subset of $\{ Y_1^{**}, \ldots , Y_{K}^{**} \}$.

The introduction of auxiliary variables leads to a refinement of the partition of the observations $\bm{Y}$ defined in Equation \eqref{eq:pEPPF}. Indeed, now we can look for the probability that the observations are partitioned into a set of $m_{\cdot, \cdot}$ distinct groups according to both tables and dishes. Such a probability coincides with an augmented version of Equation \eqref{eq:pEPPF} derived in  \citet{camerlenghi2018}, i.e., 
\begin{equation} \label{eq:pEPPF_augmented}
\begin{split}
&\Pi_k^{(n)} (\bm{n}_1, \ldots , \bm{n}_J; (n_{j,t,k})_{j,t,k}, (m_{j,k})_{j,k} )\\
& \qquad\qquad =
\Phi_K^{m_{\cdot, \cdot}} (m_{\cdot, 1}, \ldots , m_{\cdot, K}) \prod_{j=1}^J  \Phi_{m_{j, \cdot }, j}^{(n_j) } (n_{j,\cdot, 1},\dots,n_{j,\cdot, K}),
\end{split}
\end{equation}
where the functions $\Phi_K^{m_{\cdot, \cdot}}$ and $\Phi_{m_{j, \cdot }, j}^{(n_j) }$ denote the so-called \textit{exchangeable partition probability function} (EPPF) induced by $P$ and $P_j$, respectively. Then we have 
\begin{align*}
\Phi_{K}^{m_{\cdot, \cdot}} (m_{\cdot, 1}, \ldots , m_{\cdot, K}) =\frac{\prod_{i=1}^{K-1}(\theta+\sigma i)}{(\theta)_{m_{\cdot, \cdot}}}\prod_{i=1}^{K}(1-\sigma)_{m_{\cdot, i}-1},
\end{align*}
where $(a)_n: = \Gamma (n+a)/\Gamma (a)$ is the Pochhammer symbol for the rising factorial, and where $m_{j, \cdot}$ represents the total number of tables in group $j$, and $m_{\cdot,\cdot}$ is the number of tables across restaurants. An analogous explicit formula holds for the probability
$\Phi_{m_{j, \cdot }, j}^{(n_j) }$ as well. Furthermore, one can then obtain an expression
for Equation \eqref{eq:pEPPF} by integrating out the tables in Equation \eqref{eq:pEPPF_augmented}. %The dots in the indexes represent sum over that index, for example, $m_{j \cdot}$ represents the total number of clusters in group $j$, while $m_{\cdot\cdot}$ is the number of clusters across groups.

The CRF provides a simple and meaningful interpretation of the predictive distributions for observed species within and across populations. In particular, conditional on $P_j$, the predictive distribution for a new observation $Y_{j,n_j+1}$ of the $j$th population is the same as the CRP in the single population case. On the other hand, by integrating out $P_j$, we obtain the predictive distribution for the new species in population $j$ with respect to the unique species in the joint sample (across populations). That is, we can write
\begin{gather*}
    Y^*_{j,m_{j, \cdot}+1}|Y_{1,1}^*,\cdots,Y_{J,m_{J, \cdot}}^*,P \sim
\sum_{k=1}^K \frac{m_{\cdot, k}-\sigma}{\theta + m_{\cdot,\cdot}}\delta_{Y_k^{**}} +\frac{\theta + K \sigma}{\theta + m_{\cdot,\cdot}} P,
\end{gather*}
where $(Y_1^{**},\cdots,Y_K^{**})$ are the distinct species in the joint sample from $J$ populations, and $m_{j,k}$ is the number of observations in population $j$ from species $Y_k^{**}$. Notice that $m_{\cdot, k}$ is the number of times that species $Y_k^{**}$ has been observed in the joint sample.
Intuitively, a high value of $m_{\cdot, k}$ leads to a high probability of observing $Y_k^{**}$ in all populations, even if $Y_k^{**}$ has not yet been sampled in some of the $J$ populations. In particular, this probability is proportional to $m_{\cdot, k}-\sigma$: the number of times that we observe $Y_k^{**}$ minus the discount parameter of the base distribution $P$. In other words, the pair of parameters $(\theta,\sigma)$ allow us to control the total number of species in the joint sample and the extent of sharing of species across different populations. More precisely, we have that: i) if $\theta$ is low then, in expectation, the total number of distinct species in the joint sample will be low in expectation; ii) if $\sigma$ is high then, in expectation, the distinct populations will share fewer species. 

%%%%%%%%%%%%%%%%%%%%%%%%%%%%%%%%
%%%%%%%%%%%%%%%%%%%%%%%%%%%%%%%%
%%%%%%%%%%%%%%%%%%%%%%%%%%%%%%%%
%%%%%%%%%%%%%%%%%%%%%%%%%%%%%%%%

\section{The HPY-TS strategy}  \label{sec:inference}
In this section, we present our Bayesian nonparametric sequential approach, referred to as HPY-TS, for guiding the selection of samples for single cell sequencing technologies. HPY-TS is a multi-armed bandit strategy that combines the TS strategy with a Bayesian nonparametric counterpart of the GT estimator under the HPY process prior. Our HPY-TS strategy may be viewed as an extension of the strategy that has been recently proposed by \cite{battiston2016multi} from a single trial before switching arms to a pre-determined number of consecutive trials before switching arms. It also may be viewed as a Bayesian nonparametric counterpart of the GT-TS strategy proposed in \cite{dumitrascu2018}, where the smoothed GT estimator is replaced by the Bayesian nonparametric alternative under the HPY prior including a hierarchical structure on the species.

Consider $M$ cells that are simultaneously observed from multiple populations. Populations, i.e., organs, tissues, regions, or experimental conditions, represent arms to be selected for experimentation. Under the HPY prior assumption for the unknown composition of the populations, the HPY-TS strategy prescribes to select the population in such a way as to maximize the number of new distinct cell types that we expect to observe in $M$ additional cells from the selected population. We define the set of hitherto unobserved cells as $A = \{y \in \mathcal{Y}: y \notin \bm{Y} \}$, and we denote by $K_j^{(M)}| \bm{Y}$ the random number of new distinct cell types that will be observed in an additional sample of size $M$ collected from population (or arm) $j$. In such a situation the reward distribution for each arm $j$ is
the distribution of the random variable $ \mathbb{E}(K_j^{(M)} | \bm{Y})$, whose randomness is due to the fact that $P_j$ is random. We remark that, conditioning on $P_j | \bm{Y}$, then $ \mathbb{E}(K_j^{(M)} | \bm{Y})$ becomes a number. The HPY-TS strategy computes draws from the posterior distribution 
of $(\mathbb{E}(K_1^{(M)} | \bm{Y}), \ldots , \mathbb{E}(K_J^{(M)} | \bm{Y}))$, and to select the arm $j$ that corresponds to the maximum value of $ \mathbb{E}(K_j^{(M)} | \bm{Y})$. This strategy usually outperforms with respect to the so-called greedy strategy, which selects the arm with the highest posterior point estimate of $\mathbb{E}(K_j^{(M)} | \bm{Y})$. Indeed, the HPY-TS better balances the exploration step as clearly explained in \cite{battiston2016multi}.

Under the HPY process prior, $ \mathbb{E}(K_j^{(M)} | \bm{Y})$ provides the natural Bayesian nonparametric counterpart of the smoothed GT estimator proposed in \cite{dumitrascu2018}. An explicit expression for the posterior expectation $\mathbb{E}(K_j^{(M)} | \bm{Y})$ appeared in Proposition 2 of \cite{battiston2016multi}. In the next proposition, we simplify this expression. We denote by $\text{beta}(\cdot\,|\,a,b)$ the beta distribution with parameters $(a,b)$. Let $P_j$ be the unknown cell type proportions of population $j$. Let $P_j (A)$ represent the unknown cell type proportions for the collection of cells that have not yet been sampled $A$ from population $j$.

\begin{prop}\label{prop:hpy}
Let the unknown cell type proportions $P_j$ of population $j$ be modeled according to the HPY process (Equation \eqref{eq:hPY_def}). Conditional on random variables $\beta_0 | \bm{Y} \sim {\rm beta }(\beta_0|\theta+K \sigma, m_{\cdot \cdot} -\sigma K)$
 and $P_j(A)| \bm{Y}, \beta_0 = p_j$,
where
\begin{align*}
&P_j (A)|\bm{Y}, \beta_0\\
&\quad \sim{\rm beta }(p_j\,|\,(\theta_j + m_{j, \cdot}\sigma_j) \beta_0, (\theta_j + m_{j, \cdot}\sigma_j)(1-\beta_0) +n_{j, \cdot \cdot} -\sigma_j m_{j, \cdot}),
\end{align*}
one has
\begin{equation} \label{eq:EK_augmented}
\begin{split}
\mathbb{E}(K_j^{(M)} | \bm{Y}, \beta_0  , p_j) & = 
    \frac{\theta+K\sigma}{\sigma}\Big[\sum_{i=1}^{M}{M\choose i}p_{j}^{i}(1-p_{j})^{M-i}\\
    & \qquad \times\E\left[\frac{(\theta+K\sigma+\sigma)_{J_{i}}}{(\theta+K\sigma)_{J_{i}}}\right]-(1 - (1 - p_j)^M)\Big]
\end{split}
\end{equation}
with 
\begin{align*}
    \E\left[\frac{(\theta+K\sigma+\sigma)_{J_{i}}}{(\theta+K\sigma)_{J_{i}}}\right] = \sum_{\tilde{m}=1}^{i}F(i,\tilde{m},\sigma,(\theta+m_{j\cdot})\beta_{0})\frac{(\theta+K\sigma+\sigma)_{\tilde{m}}}{(\theta+K\sigma)_{\tilde{m}}}.
\end{align*}
Here, random variable $J_i$, for any $i=1,\ldots,M$, counts the number of distinct values in a random sample of size $i$ from a PY process with updated parameters $(\sigma,(\theta+m_{j, \cdot})\beta_{0})$, and $F(n,k,\sigma,\theta)$ is the probability that $\{J_i = \tilde{m}\}$.
\end{prop}

The proof of Proposition \ref{prop:hpy} is deferred to the Supplementary Material. Based on Proposition~\ref{prop:hpy}, one can recover an explicit formula for $\E(K_j^{(M)} | \bm{Y})$ by simply integrating Equation \eqref{eq:EK_augmented} with respect to the distribution of $p_j$ and the distribution of $\beta_0$. Then, we can infer that the computational complexity of computing the formula for $\E(K_j^{(M)} | \bm{Y}, \beta_0, p_j)$ is proportional to $M$. Having found the posterior expectation of $K_j^{(M)}|\bm{Y}$ for all populations $j$ in Proposition \ref{prop:hpy}, our HPY-TS strategy selects the population with the highest expected rewards, computed from a posterior sample. More specifically, we sample $\beta_0 | \bm{Y}$ and $P_j(A)| \bm{Y}, \beta_0$ from the distribution described in Proposition \ref{prop:hpy}. Then, conditional on these realizations, we compute $\E(K_j^{(M)} | \bm{Y}, \beta_0, p_j)$ according to Equation \eqref{eq:EK_augmented}. Finally, we select the population with the highest realized value. Details of the HPY-TS strategy are described in Section \ref{sec:algo}. Our HPY-TS strategy is based on the Thompson's sampling approach (Algorithm \ref{algo:TS}), with parameters updated sequentially according to Algorithm \ref{algo:parameters}.

With regards to the choice of the prior distributions for the hyperparameters of the HPY process, we assume a Uniform prior on $(0,1)$ for both parameters $\sigma$ and $\sigma_j$. Moreover, we assume a Gamma prior with parameters $(1,1)$ for both parameters $\theta$ and $\theta_j$. All prior distributions are assumed to be independent. Note that Algorithm \ref{algo:TS} depends on the following collection of parameters
\begin{displaymath}
\eta = (\theta,\sigma,\sigma_j,\theta_j;\, j = 1,\cdots, J),
\end{displaymath}
and on the table counts of the CRF, which are encoded by the vector $\bm{m}_J=(m_{j,\cdot}; \,  j = 1,\cdots, J)$. Here it is worth stressing that the collection of table counts $\bm{m}_J$ are latent variables, that is quantities that have not been observed in the initial sample. Therefore, before running Algorithm \ref{algo:TS}, we estimate these latent variables. This is done by using a Gibbs sampling algorithm that relies on the explicit expression of the pEPPF from the work of \cite{camerlenghi2018}. Specifically, we exploit the sequential structure of our species sampling problem to update the vector of parameters $\eta$: we describe the novel and efficient algorithm for the updating of $\eta$ in the next section. 

%%%%%%%%%%%%%%%%%%%%%%%%%%%%%%%%
%%%%%%%%%%%%%%%%%%%%%%%%%%%%%%%%
%%%%%%%%%%%%%%%%%%%%%%%%%%%%%%%%
%%%%%%%%%%%%%%%%%%%%%%%%%%%%%%%%

\section{Sequential parameter updates} \label{sec:algo}

The multi-armed bandit problem is described as a sequential allocation problem, where the goal is to find the best allocation strategy to sample new observations from $J$ different populations at every experimental time step. Whenever the new $M$ cells are sampled from a population, one has to update the parameters of the HPY process in a computationally feasible way. A possible approach to this problem was first suggested in \cite{battiston2016multi}, where the authors propose a Markov chain Monte Carlo (MCMC) in order to estimate the posterior distributions of the hyperparameters of the HPY. However, such an approach does not take advantage of the sequential nature of the species sampling problem and, more importantly, is not computationally feasible with large data sets. The computational burden of the approach of \citet{battiston2016multi} makes its direct application almost impossible, except for toy examples with small numbers of arms. In this section, we suggest a computationally tractable approach that leverages the sequential structure of the problem (Algorithm 1) and is based on a filtering algorithm of \cite{liu2001combined}.

\begin{algorithm}
\caption{HPY-TS}
\begin{algorithmic} \label{algo:TS}
\vspace{0.4cm}
 \FOR{$i \in 1$:number of new samples}
\vspace{0.1cm}

\STATE {draw $\beta_0 \sim {\rm beta}(\theta + \sigma K ,m_{\cdot ,\cdot} -  \sigma K)$
\vspace{-0.1cm}

\STATE \FOR{$j \in 1:J$}
\vspace{0.1cm}
\STATE {draw $p_j \sim {\rm beta} ((\theta_j + m_{j,\cdot}\sigma_j)\beta_0,(\theta_j + m_{j,\cdot}\sigma_j)(1-\beta_0) + n_{j, \cdot, \cdot } - \sigma_j m_{j ,\cdot})$ \\ 
Compute $ \E(K_j^{(M)} | \bm{Y}, \beta_0, p_j)$ according to Proposition \ref{prop:hpy}}
\vspace{0.1cm}
\ENDFOR \\
\vspace{0.1cm}
Compute $j^*  = argmax \{ \E(K_j^{(M)} | \bm{Y}, \beta_0 , p_j), \; j = 1,\cdots, J\}$; \\
\vspace{0.1cm}
Draw the next sample from population $j^*$; \\
\vspace{0.1cm}
Update the HPY parameters according to Algorithm \ref{algo:parameters};
\vspace{0.1cm}
}
\ENDFOR 

\vspace{0.4cm}
\end{algorithmic}
\end{algorithm}

The HPY-TS strategy selects the arm to sample from, then one sequentially samples the batch of $M$ cells from the selected 
arm. After that, one updates the model parameters with the new observation encoded by $\eta$ to select the new arm to sample from. We then consider discrete time points $t =1, 2, \ldots$, and we clarify how to sequentially update the parameters of our model in a computationally feasible way. Let us fix some notation: $y_t$ is the vector of observations from the arm selected at time $t$, and $D_t=\{y_t,D_{t-1}\}$ is the set of observations available at time $t$. Thus, we can think of a model that is described by a distribution $p(y_t | \eta)$ evolving in time and depending on a vector of model parameters $\eta$. At each iteration, we select an arm and observe $y_{t+1}$, and we sample the updated parameters from the posterior distribution $p( \eta | D_{t+1})$, as $t=1, 2, \ldots$. Note that this posterior distribution is proportional to 
\begin{displaymath}
p(y_{t+1} | \eta, D_t) p(\eta | D_{t}),
\end{displaymath}
due to Bayes' theorem. We can think of $p(\eta | D_{t})$ as the density function of $\eta$ at time $t$. Our aim is to sample a new set of parameters from the posterior distribution of $\eta$ in the presence of a new observation $y_{t+1}$. The key idea is to approximate the distribution of $\eta| D_t$ with a mixture of $N$ Gaussian kernels, i.e.,
\begin{align*}
p(\eta | D_{t}) \approx \sum_{i=1}^N \omega_t^{(i)} {\mathcal N} (\eta | m_t^{(i)},h^2 V_t),
\end{align*}
where $\{\omega_{t}^{(1)},\dots,\omega_{t}^{(N)}\}$ is the set of importance sampling weights for $\{\eta_t^{(i)}: i=1,2,\dots,N \}$ at time $t$, $N$ is the number of importance samples at each time step, and ${\mathcal N} (\eta |m, V)$ is the density function of a Gaussian distribution with mean $m$ and covariance $V$. Moreover, $V_t$ is the estimate of the covariance with respect to the Monte Carlo posterior, and $h$ is a smoothing parameter. \cite{liu2001combined} suggest to choose $h$ as a decreasing function of the number of importance samples. In our simulations, we set $h = \frac{1}{N}$. In order to avoid ``loss of information'' over time, earlier work~\citep{west1993a,west1993b} proposes shrinkage kernel locations and suggests setting $m_t^{(i)}=a\eta_t^{(i)}+(1-a)\bar{\eta}_t$ and $a = \sqrt{1 - h^2}$, where $\bar{\eta}_t$ is the mean of the Monte Carlo sample of size $N$ at time $t$. With these choices, we preserve the covariance $V_t$ over time.

\begin{algorithm}
\caption{Filtering algorithm}
\begin{algorithmic} \label{algo:parameters}
\vspace{0.4cm}
\STATE Evaluate $m_t^{(i)}$ for each $i=1,2,\ldots,N$:
\begin{align*}
m_t^{(i)}=a\eta_t^{(i)}+(1-a)\bar{\eta}_t,
\end{align*}
which are the prior point estimates of $\eta$. Construct a posterior approximation of $p(\Theta | D_{t+1})$ with weights $\omega_{t+1}^{(i)}$ and samples $\eta_{t+1}^{(i)}$, for $i=1, \ldots , N$, as follows.
\STATE \FOR{$i \in 1:N$}
\vspace{0.1cm}
\STATE  \begin{itemize}
\item[1)] Sample an auxiliary integer variable $k$ from the set $\{1,2,\ldots,N\}$ with probability proportional to:
\begin{align*}
g_{t+1}^{(i)} \propto \omega_t^{(i)} p(y_{t+1} | m_t^{(i)}, D_t)
\end{align*}
\item[2)] Sample a new parameter vector $\eta^{(k)}_{t+1}$ from the $k$th normal component of the kernel density, namely:
\begin{align*}
\eta_{t+1}^{(k)} \sim N(m_t^{(k)},h^2 \bm{V}_t)
\end{align*}
\item[3)] Evaluate the corresponding weights
\begin{align*}
\omega_{t+1}^{(k)} \propto \frac{p(y_{t+1} | \eta_{t+1}^{(k)},  D_t)}{p(y_{t+1} | m_{t}^{(k)},  D_t)}
\end{align*}
where
\begin{align*}
p(y_{t+1} | \eta,  D_t)\propto %EPPF(m_{t, 1},\cdots,m_{t, K} | \alpha_{t+1}^{(k)},\gamma_{t+1}^{(k)}) \prod_{j =1}^J EPPF(n_{j \cdot 1},\cdots,n_{j \cdot k_j} | \cdot,\sigma_{j,t+1}^{(k)})
\Pi_k^{(n)} (\bm{n}_1, \ldots , \bm{n}_J; (n_{j,t,k})_{j,t,k}, (m_{j,k})_{j,k} ).
\end{align*}
In other words $p(y_{t+1} | \eta,  D_t)$ is proportional to the pEPPF defined in \eqref{eq:pEPPF_augmented}, depending on the parameters $\eta$ and on information available up to time $t+1$.
\end{itemize}
\ENDFOR 

\vspace{0.1cm}

\STATE Resample according to the importance weights $\omega_{t+1}^{(k)}$ to obtain a set of parameters with equal weights--in other words, a Monte Carlo approximation of the posterior.

\vspace{0.4cm}
\end{algorithmic}
\end{algorithm}

In our framework, one only needs to evaluate the conditional distribution $p(y_{t+1} | \eta, D_t)$, which may be recovered from the expression of the pEPPF in Equation \eqref{eq:pEPPF_augmented}. In particular, we initially run a Gibbs sampler \citep{camerlenghi2018} to obtain a collection of random samples for the latent table counts and parameters $\eta$. Then, the output of the initial sample may be regarded as an importance sample of $\eta$ with equal weights at time $t=0$. 
More precisely, we used the Gibbs sampling scheme  described in \cite{camerlenghi2018} to initialize the values of all of the parameters $\eta$, conditionally on the data $D_0$, which contains the observations $\bm{Y}$. In fact, we run the MCMC procedure of \cite{camerlenghi2018}, and we used the output of the last $N$ runs to initialize all the parameters for the particle filtering algorithm. We also assign uniform importance sampling weights to all of the initial particles, i.e., $\omega_0^{(i)}=1/N$ as $i=1, \ldots , N$.
Then, we use Algorithm \ref{algo:parameters} to sequentially update the parameters of the HPY.
 
%%%%%%%%%%%%%%%%%%%%%%%%%%%%%%%%
%%%%%%%%%%%%%%%%%%%%%%%%%%%%%%%%
%%%%%%%%%%%%%%%%%%%%%%%%%%%%%%%%
%%%%%%%%%%%%%%%%%%%%%%%%%%%%%%%%

\section{Applications} \label{sec:application} 

\subsection{Simulation study} \label{sec:simulations}
We first demonstrate the performance of our HPY-TS algorithm in the context of simulated data.
Additional simulation studies are presented in Section \ref{sec:add_simulations} of the Supplementary Material. We consider a setup with $100$ arms, representing a sample corresponding to $20000$ different species. The true distribution of each arm follows Zipf's law, such that the mass assigned to the $k^{th}$ most common species in a population $j$ is
\[
p_j(k;s_j)=\frac{1/k^{s_j}}{\sum_{i=1}^{N_j} 1/i^{s_j}},
\]
where $N_j>0$ is the number of species in population $j$, and $s_j>1$ is a real parameter that controls the distribution of mass among the support--a large $s_j$ indicates that the total mass is concentrated on a few points, and a small value indicates that the mass is shared across many points. Hence, an arm with a low $s_j$ is a `winning arm,' or an arm with high species diversity. Among the $100$ arms, we consider $4$ winning arms (Zipf parameter $s_j=1.3$), and $96$ less diverse arms (Zipf parameter $s_j=2$). An optimal strategy should balance exploration and exploitation, and query the less diverse arms occasionally, while focusing on the winning arms.  

We evaluate the performance of our TS strategy by comparing it with three baselines: the Oracle strategy, the Uniform strategy, and the Good-Toulmin Thompson sampling (GT-TS) strategy proposed by \cite{dumitrascu2018gt}. In particular, the Oracle strategy is used to compare the performance with the optimal behavior; the Oracle strategy is allowed to see into the future or pre-sample from all the arms and make the optimal decision at every iteration.
More precisely, the Oracle strategy selects the arm having the highest probability of observing a new cell, where such a probability is evaluated numerically assuming knowledge of the true distribution of the data (i.e., a Zipf distribution in our experiments).
The Uniform strategy allocates the budget uniformly across the arms, whereas the GT-TS strategy is based on a smoothed version of the Good-Toulmin estimator. More precisely, the smoothed Good-Toulmin estimator~\citep{orlitsky2016optimal} estimates the number of new species that will be sampled in an additional sample of size $M$ for a fixed population $j$. This estimator is defined as
\[
    \widehat{U}_j^{(M)}(\bm{Y}_j) = - \sum_{i=1}^\infty (- M/n_j)^i \mathbb{P}(L > c_j)\Phi_{i_j},
\]
where $M/n_j$ is referred to as the \emph{extrapolation factor}, $\Phi_{i_j}$ denotes the number of species occurring with frequency $i$ in $\bm{Y}_j$, the random  sample from the $j$th population, and $L$ is an independent random nonnegative integer. Common choices for the distribution of the random variable $L$ include the Poisson distribution and the binomial distribution \citep{orlitsky2016optimal}. The Good-Toulmin diversity estimator can be incorporated into the multi-armed bandit framework as follows. At each sampling step, an arm is chosen based on its probability of yielding the greatest number of novel species. The probability that the $j$th population is chosen during a trial is based on the weight of its Good-Toulmin estimator $\widehat{U}_j^{(M)}(\bm{Y}_j)$. Upon collecting $M$ new samples from the chosen arm, the reward (the number of novel cell types) is observed, and the parameters of the Good-Toulmin estimator for the chosen population are reestimated with the new samples and reward \citep{dumitrascu2018gt}.

\begin{figure*}[h!]
\includegraphics[width=0.7\linewidth,height=7cm]{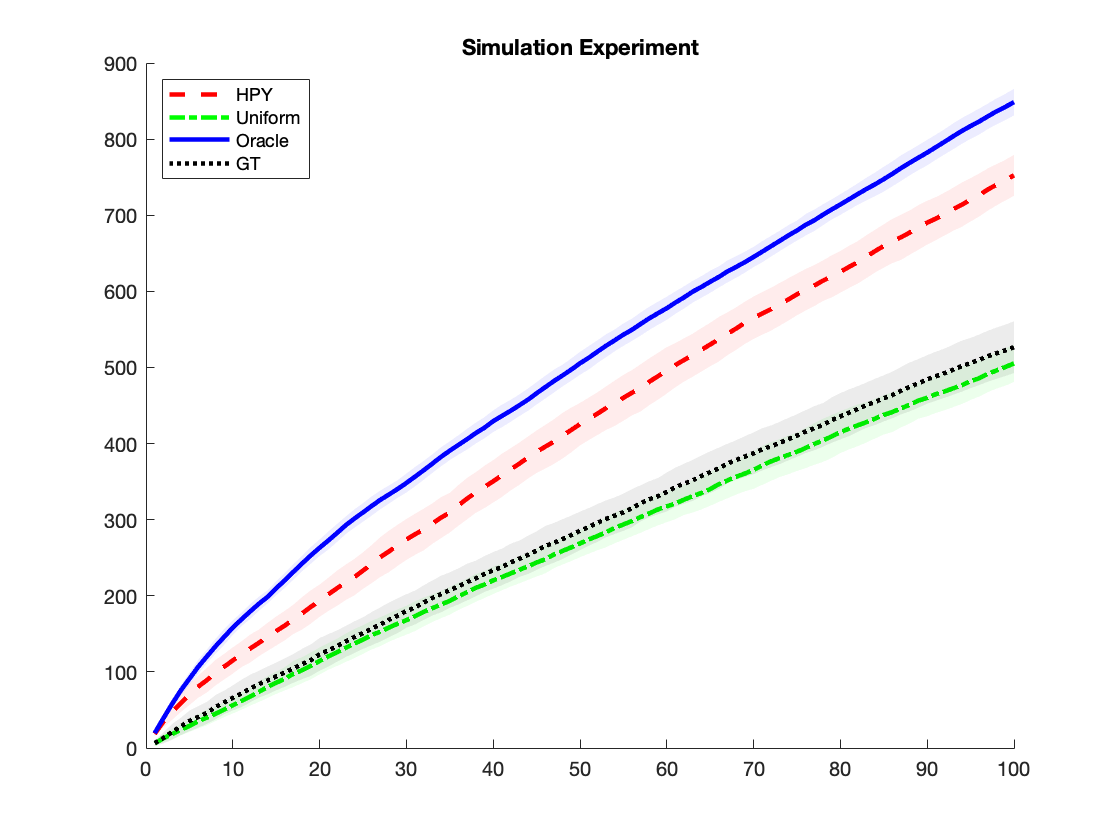}
    \caption{\textbf{Simulation results.} We consider a multi-armed bandit setting for population sampling with $100$ arms in which the species diversity follows Zipf's law with parameters $1.3$ ($4$ high diversity, winning arms) and $2$ ($96$ low diversity arms). An initial sample of $20$ cells were collected from each of the $100$ arms, with $50$ additional cells sampled at each iteration. We used $100$ sampling steps and averaged the results over $50$ runs. We compared HPY-TS (red, dashed) to two baselines---the GT-TS sampler (black, dotted) and a Uniform sampling strategy (green, dot-dashed line)---and to an Oracle estimator (blue, solid). The shaded bands are within one standard deviation of the average performance, computed as the mean across simulations.}\label{fig_100}
\end{figure*}

In implementing our HPY-TS strategy, we use an initial sample of $20$ observations from each of the $100$ arms, with $M=50$ observations sampled at each iteration. We use $500$ sampling steps, and the results are averaged over $50$ runs. The computations are performed in parallel, and the code is available at \url{https://github.com/fedfer/HPYsinglecell}. We observe that the HPY-TS algorithm performs better than the GT-TS strategy and the Uniform strategy; the latter two methods explore, but fail to exploit the most diverse arms (Fig. \ref{fig_100}). As expected, the HPY algorithm discovers fewer new species than the Oracle strategy, but the HPY approach comes close to Oracle behavior. The results show similarities with the performance previously reported in the work of \cite{battiston2016multi} for a simulation scenario with a small number of arms ($8$ arms), with the added benefit of \emph{scalability} to an order of magnitude more arms. The good performance of the algorithm has been assessed in Section \ref{sec:diagnostics} of the Supplementary Material through posterior diagnostics for  the simulation scenario considered in this section. 

\subsection{Application to single cell RNA-seq experimental design}  \label{sec:numerical} 

\begin{figure*}[h!]
    \includegraphics[width=\linewidth]{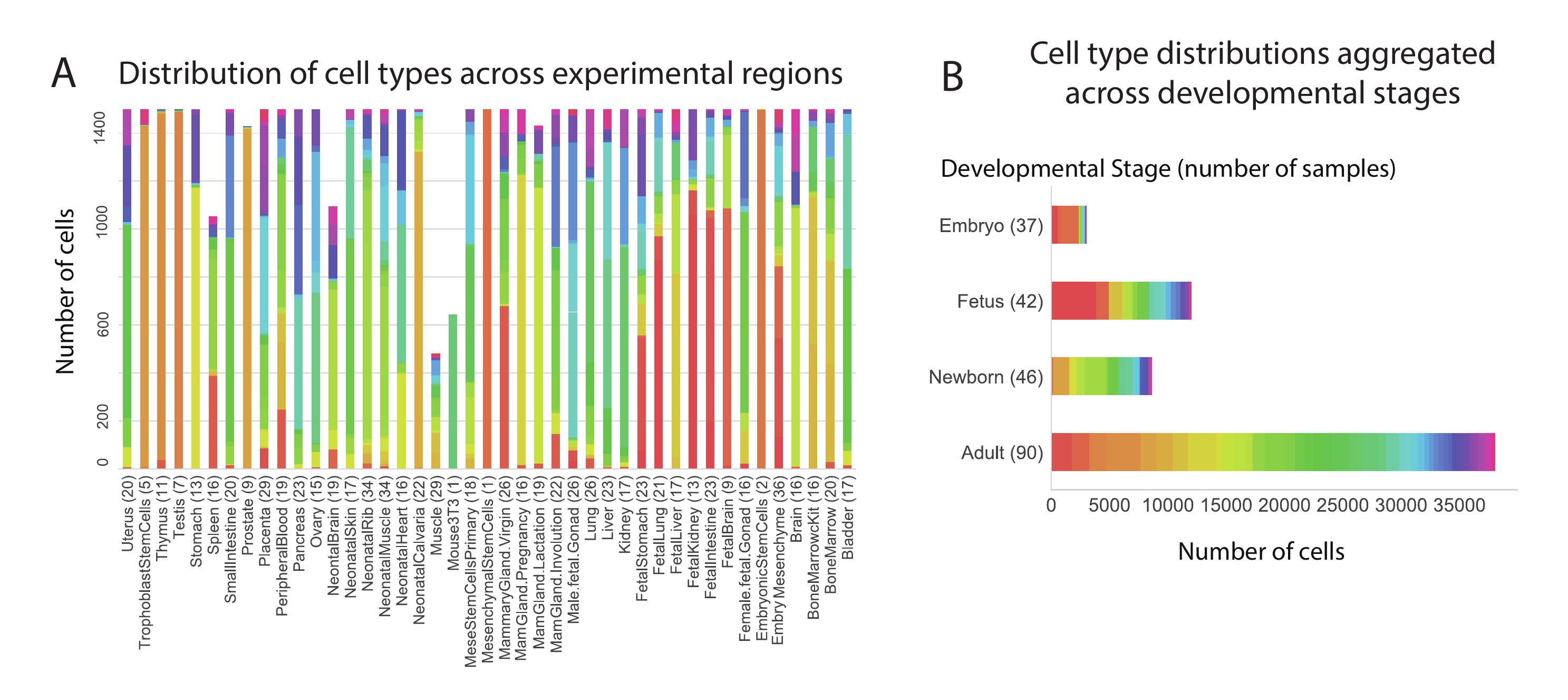}
    \caption{Summary figure from \citet{dumitrascu2018gt} of the single cell RNA-seq data from the Mouse Cell Atlas \citep{han2018mapping}. The different colors represent different cell types. 
    Panel A: Cell type distributions across tissues
    together with the corresponding cells and specimens. Panel B: Cell type distributions per arm: aggregated tissue types and developmental stages.}\label{fig0}
\end{figure*}

We further illustrate the advantage of our approach in the context of a simulation study based on the Mouse Cell Atlas data \citep{han2018mapping}.
The Mouse Cell Atlas aims to provide the first high-throughout transcriptome-based single-cell atlas in a mammalian system. The project assayed over $400,000$ cells from all of the major mouse organs and identified previously uncharacterized cell populations (Fig.~\ref{fig0}). Following technical noise correction, $60,000$ high-quality cells were sequenced, representing $43$ distinct tissues and $98$ major cell types across four developmental stages -- embryo, fetal, newborn, and adult. In the collection process, equal numbers of cells were sampled uniformly across organs and developmental stages. We show that our experimental design approach achieves similar cell type diversity while requiring substantially fewer samples when compared to related methods. We follow the simulation setup developed in prior work \citep{dumitrascu2018gt}, outlined below.

In our simulation study, we envision a setting in which the cells were assayed in smaller batches than in the actual experiments. In particular, smaller batches are common in single cell experiments that use technologies that are less noisy but more expensive; thus experimental design plays an important role in minimizing cost \citep{angerer2017single}. Moreover, larger batches would quickly saturate the available data, so we evaluate on batch sizes that are typically smaller than are used. The $43$ mouse organs were aggregated across the four developmental stages---embryo, fetal, newborn, and adult---resulting in a heterogeneous data set. Cells were sampled with replacement from each of the four experimental categories (arms), representing the four developmental stages. An experimental round corresponds to an allocation step in which the cell budget is distributed across the four experimental conditions. 

\begin{figure}[h!]
\includegraphics[width=0.9\linewidth,height=8cm]{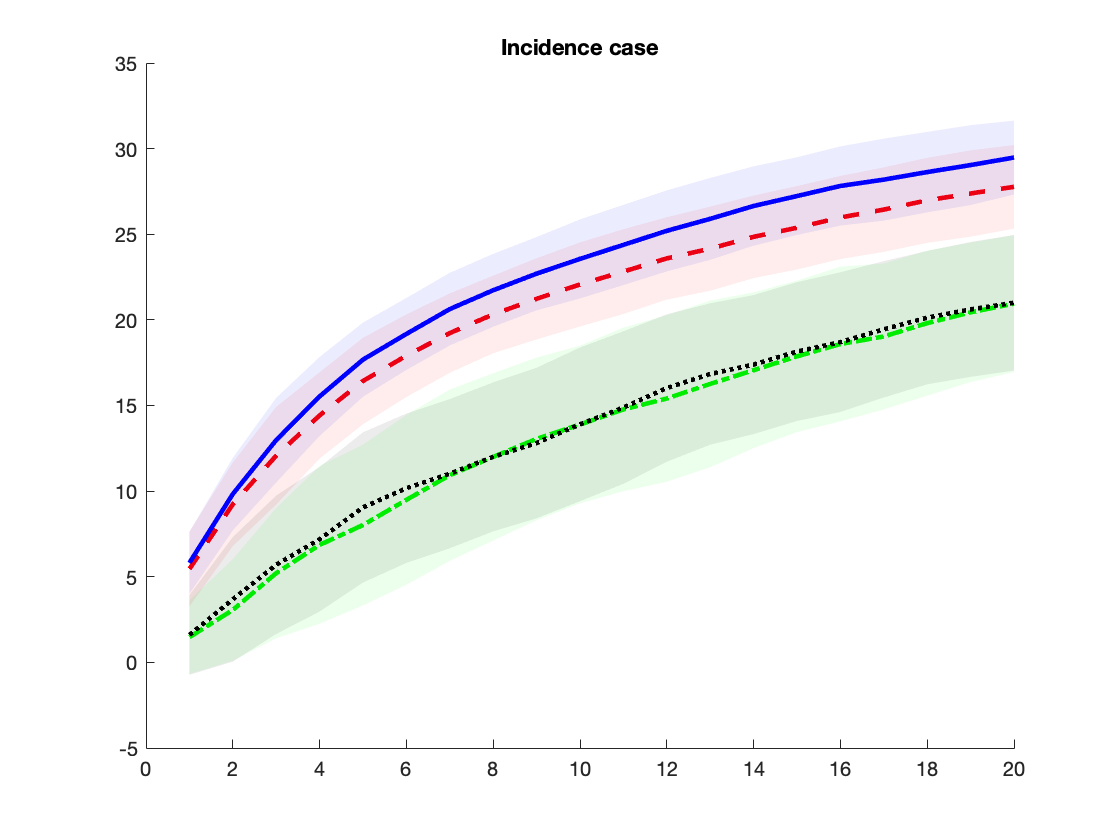}
\caption{\textbf{Performance of HPY-TS on the Mouse Cell Atlas data (incidence case).} An initial sample of $M=50$ cells were collected from each of four populations: \textit{embryo, fetal, newborn} and \textit{adult}. Following the initial sampling step with $M=50$ samples, $20$ sequential trials were performed. At each time step, all $M=25$ cells were sampled from one chosen experiment. The results were averaged over $100$ runs of each algorithm. We compared HPY-TS (red, dashed) to two baselines---the GT-TS sampler (black, dotted) and a Uniform sampling strategy (green, dot-dashed line)---and to an Oracle estimator (blue, solid). The shaded bands are within one standard deviation of the average performance, computed as the mean across simulations.}
\label{Fig1}
\end{figure}

\begin{figure}[h!]
\includegraphics[width=0.9\linewidth,height=8cm]{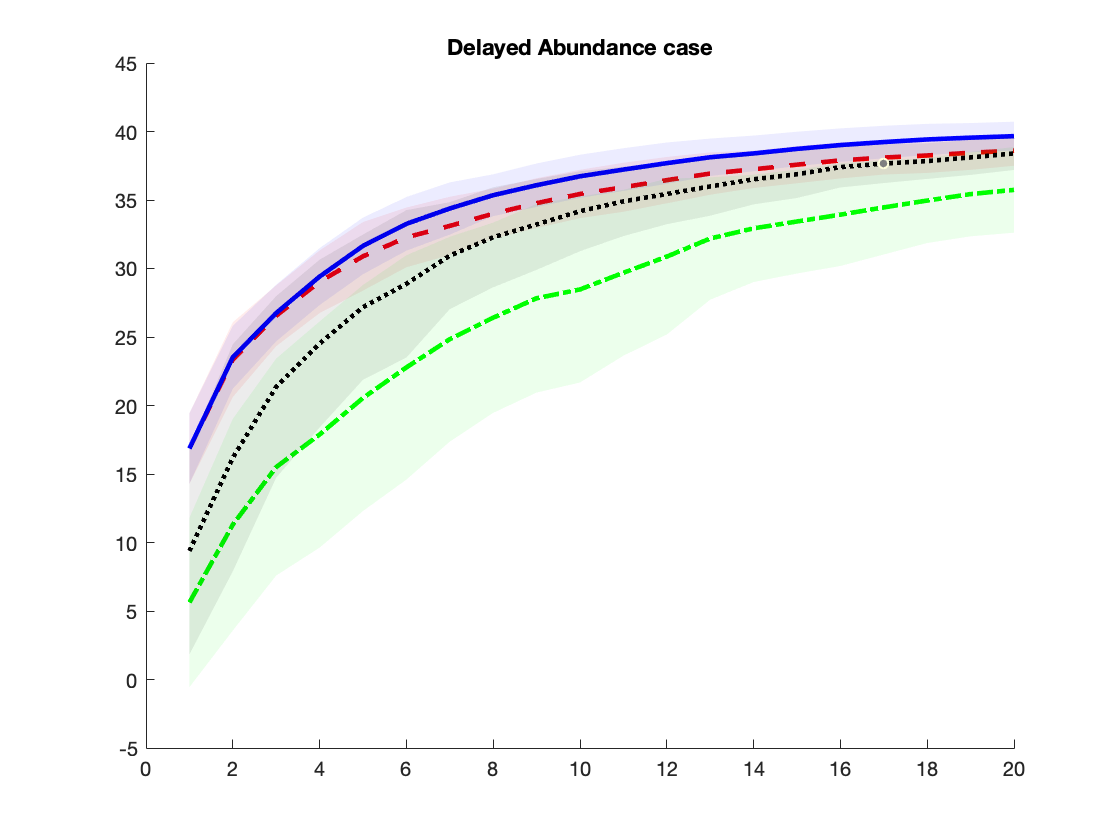}
\caption{\textbf{Performance of HPY-TS on the Mouse Cell Atlas data (delayed abundance).} An initial sample of $M=50$ cells were collected from the four populations: \textit{embryo, fetal, newborn} and \textit{adult}. Following the initial sampling step, $20$ additional trials were performed. At each time step, $M=100$ samples were distributed across the arms following a diversity estimation step. The results were averaged over $100$ runs of each algorithm. We compared HPY-TS (red, dashed) to two baselines---the GT-TS sampler (black, dotted) and a Uniform sampling strategy (green, dot-dashed line)---and to an Oracle estimator (blue, solid). The shaded bands are within one standard deviation of the average performance, computed as the mean across simulations.}
\label{Fig2-delayed}
\end{figure}

We consider two ways of allocating samples: the \emph{incidence case} and the \emph{delayed abundance} case. In the incidence case (see Fig. \ref{Fig1}), a single most informative experimental condition is chosen, and $M$ samples come from that single condition. In the delayed abundance case (see Fig. \ref{Fig2-delayed}), samples are allocated across all of the available experimental conditions in parallel. In both cases, the budget allocation step is applied using the HPY-TS strategy as follows. In the incidence case, we allocate more cells to the experiment (i.e., developmental stage) with a higher probability of yielding new cell types based on previous trials. Following the initial sampling step with $M=50$ samples from each arm, $20$ additional trials were performed. At each time step, all $M=25$ cells were sampled from one chosen experiment. In the delayed abundance case, after the initial $M=50$ samples from each arm, a budget of $M=100$ cells were distributed across arms according to the HPY-TS estimated probabilities, across $20$ sequential trials.  The results were averaged over $100$ runs for each algorithm, and the HPY-TS sequential Monte Carlo strategy used $500$ sampling steps. We compare HPY-TS to three other approaches--the GT-TS sampler, a Uniform sampling strategy, and an Oracle estimator.

Our results show that the HPY-TS approach achieves substantial improvement in efficiency as compared to the baseline GT-TS estimator and to the Uniform sampling strategy (Fig.~\ref{Fig1}).  Moreover, the HPY-TS approach shows nearly optimal performance, as compared with the performance of the Oracle strategy. When compared to the Uniform strategy, our HPY-TS approach leads to, on average, as much as $50\%$ more distinct cell types identified, with an average consistent margin of $10$ additional distinct cell types identified across trials (Fig. \ref{Fig1}, \ref{Fig2-delayed}).  The baseline GT-TS approach approximates the probability of observing a new cell type according to a model that assumes the cell types are distributed according to a Poisson process \citep{orlitsky2016optimal}. In contrast, the HPY-TS algorithm assumes that all arms share a baseline distribution given by the base measure, information that is diffused across the developmental landscape to generate the developmental stage-specific cell type distributions. Sharing information across experiments using this prior appears to substantially improve performance by allowing updates of the parameters governing experiments similar the chosen experiment at each iteration, instead of only updating the chosen experiment's parameters.

%%%%%%%%%%%%%%%%%%%%%%%%%%%%%%%%
%%%%%%%%%%%%%%%%%%%%%%%%%%%%%%%%
%%%%%%%%%%%%%%%%%%%%%%%%%%%%%%%%
%%%%%%%%%%%%%%%%%%%%%%%%%%%%%%%%

\section{Discussion}  \label{sec:discussion}
We propose the HPY-TS multi-armed bandit strategy, which uses the Thompson sampling strategy and a hierarchical Pitman-Yor process prior to optimize species discovery in experimental design. The HPY-TS strategy was shown to substantially improve cell type discovery in the setting of experimental design for single cell sequencing experiments. In particular, the HPY-TS strategy readily applies to cases where the number of arms corresponding to experimental conditions have substantial structure across those conditions. In particular, as cell atlases emerge, the strategy developed here is crucial to efficiently and effectively study cell type variability across new and growing experimental conditions including many thousands of simultaneous cellular perturbations (e.g., Perturb-seq \citep{dixit2016perturb}) and combinatorial interventions \citep{horlbeck2018mapping}. The improvements that the HPY-TS strategy achieves over uniform experimental design strategies in both simulated and real data justify incorporating these types of methods in the data collection pipeline during the experimental process.

From a statistical standpoint, our work proposed a sequential Monte Carlo scheme that, unlike the previous work of \cite{battiston2016multi}, scales to a multi-sample setting and allows for inference across a large number of experiments as one finds in cell atlas development or Perturb-seq experiments. This makes our HPY-TS strategy appropriate for experimental setups with a large and growing number of arms.  In this paper we demonstrate a number of advantages in using Bayesian experimental design to maximize cell type discovery within a budget during single cell RNA-sequencing experiments. We further show evidence that modeling the cell type structure of single cell data using an HPY prior captures the developmental constraints guiding cell type diversity and allows each sample to inform all of the arms, leading to near-Oracle behavior. 

As mentioned in Section \ref{sec:numerical}, the optimization of cell type discovery in a multi-tissue setting was proposed in \cite{dumitrascu2018gt}, which uses a strategy based on the Good-Toulmin estimator. However, the approach of \cite{dumitrascu2018gt} is empirical rather than model-based. Indeed, some paramount statistical challenges remain unsolved: i) how to model both dependence and heterogeneity across tissues in a principled statistical way? ii) how to incorporate uncertainty quantification across experimental conditions (arms) to guide arm selection at each step? iii) does a suitable statistical model, answering i)--ii), fundamentally improve performance?
Our Bayesian nonparametric approach takes into account all these challenges: the hierarchical structure allows information to be shared across similar tissues or experimental conditions, and the uncertainty in estimated rewards across experimental conditions is incorporated in the strategy through fast posterior computations (see Proposition \ref{prop:hpy}). Finally, the experiments in Section 5 and Section 2 of the Supplementary Material show that our model, which answers i)--ii), is able to discover more cell types sampling fewer cells with respect to the competing strategies.

In conclusion, the proposed HPY-TS strategy outperforms the current state-of-the-art strategies, and our contribution paves the way for future research in the field. We first emphasize that the number of cells that can be collected per trial $M$ has been assumed to be fixed, since this is typically the case in cell experiments. A possible alternative, which we do not consider in our paper, would focus on optimizing each time for how many samples $M$ should be collected over a total fixed available budget. Secondly, a natural question stemming from our analysis is understanding the effect that batch correction and cell type matching have on optimal budget allocation. In order to distinguish new cell types in a true online fashion, appropriate algorithms are needed to cluster data from new experiments, as well as reconcile the identified clusters with previously discovered ones (batch correction), in the likely presence of experiment-specific noise. In this paper, we focus on optimal experimental design under the assumption that a precise label is available at the time of the experiment. Understanding the effect of a suboptimal, possibly incorrect, or time-delayed label has on optimal experimental design is an additional area of focus for future work.

\section*{Acknowledgements}
The authors would like to thank the Associate Editor and the Referees for their valuable comments and suggestions which led to a substantial improvement of the paper.
The authors are indebted to David B Dunson, Jordan Bryan for helpful comments. Federico Camerlenghi and Stefano Favaro received funding from the European Research Council (ERC) under the European Union's Horizon 2020 research and innovation programme under grant agreement No 817257. Federico Camerlenghi and Stefano Favaro gratefully acknowledge the financial support from the Italian Ministry of Education, University and Research (MIUR), ``Dipartimenti di Eccellenza" grant 2018-2022. Bianca Dumitrascu is supported by the National Science Foundation (NSF) under grant DMS-1638352 and completed part of this work at Princeton University and while visiting The Statistical and Applied Mathematical Sciences Institute in Durham, NC, under the kind support of the NSF grant DMS-1638521. Federico Ferrari is partially supported by grant 1R01ES028804-01 of the National Institute of Environmental Health Sciences of the United States Institutes of Health. Barbara E Engelhardt is supported by NIH R01 HL133218 and an NSF CAREER IIS-1750729.

\section*{Conflict of Interest}
Barbara E Engelhardt is on the SAB for Celsius Therapeutics and Freenome, and is currently
employed by Genomics plc and Freenome and on a year leave-of-absence
from Princeton University.

%%%%%%%%%%%%%%%%%%%%%%%%%%%%%%%%%%%%%%%%%%%%%%
\begin{supplement}
\textbf{Supplementary material to ``Nonparametric Bayesian multi-armed bandits for single cell experiment design''}.
\slink[doi]{}
\sdescription{The supplementary material contains proofs, posterior diagnostics. and additional simulation studies.}
\end{supplement}
%%%%%%%%%%%%%%%%%%%%%%%%%%%%%%%%
%%%%%%%%%%%%%%%%%%%%%%%%%%%%%%%%
%%%%%%%%%%%%%%%%%%%%%%%%%%%%%%%%
%%%%%%%%%%%%%%%%%%%%%%%%%%%%%%%%

\bibliographystyle{imsart-nameyear}
\bibliography{ref.bib}

\end{document}